\documentclass{mn2e}

\usepackage{graphics}

\begin{document}

\title{H$\alpha$ kinematics of a $z\sim 1$ disc galaxy from near-IR integral field spectroscopy}

\author[J.K.~Smith et al.]{J.K.~Smith,$^{1}$ \thanks{joanna@ast.cam.ac.uk} A.J.~Bunker,$^{1,2}$ N.P.~Vogt,$^{3}$ R.G.~Abraham,$^{4}$ A.~Arag\'{o}n-Salamanca,$^{5}$ \newauthor R.G.~Bower,$^{6}$ I.R.~Parry,$^{1}$ R.G.~Sharp$^{1,7}$ and A.M.~Swinbank$^{6}$ \\ $^{1}$\,Institute of Astronomy, University of Cambridge, Madingley Road, Cambridge, CB3\,0HA, UK\\ $^{2}$\,School of Physics, University of Exeter, Stocker Road, Exeter, EX4 4QL, UK \\ $^{3}$\,Department of Astronomy, New Mexico State University, Las Cruces, NM 88003-8001, USA\\ $^{4}$\,Department of Astronomy and Astrophysics, University of Toronto, 60 St. George Street, Toronto, ON M5S 3H8, Canada \\ $^{5}$\,School of Physics and Astronomy, University of Nottingham, NG7 2RD, UK \\ $^{6}$\,Department of Physics, University of Durham, South Road, Durham DH1 3LE, UK\\ $^{7}$\,Anglo-Australian Observatory, PO Box 296, Epping, NSW 1710, Australia}

\maketitle

\begin{abstract} 
In this letter we present the first 3D spectroscopic study of
H$\alpha$ emission in a $z\sim1$ field galaxy with an integral field
unit. Using the CIRPASS spectrograph on Gemini-South we map the
spatial and velocity distribution of H$\alpha$ emission in the
$z=0.819$ galaxy CFRS 22.1313. We detect two H$\alpha$ emitting
regions with a velocity separation of 220\,$\pm$\,10 km
s$^{-1}$. Combining the 2D map of H$\alpha$ emission with HST F814W
imaging, we determine a lower limit of 180\,$\pm$\,20 km s$^{-1}$ for
the rotation velocity of this $M_B$(rest)\,$\sim$\,-21 galaxy. We note
that our value is significantly higher than the rotation velocity of
120\,$\pm$\,10km s$^{-1}$ reported by Barden et al. (2003) for their
long-slit spectroscopic study of this galaxy. Our lower limit
on the rotation velocity is entirely consistent with no evolution of
the rest $B$-band Tully-Fisher relation. The position of this galaxy
relative to the mean rest $B$-band Tully-Fisher relation of Tully \&
Pierce (2000) is consistent with brightening of no more than $\sim$\,1\,mag at
$z=0.8$. A larger integral field unit sample, without the
uncertainties inherent to long-slit samples, is needed to accurately
determine the evolution of the Tully-Fisher relation out to $z\sim1$.

\end{abstract}


\section{Introduction}
The Tully-Fisher (TF) relation describes the strong correlation
between luminosity and maximum rotation velocity for disc galaxies
(Tully \& Fisher 1977). This reflects a fundamental relationship
between the total mass of the galaxy and the mass contained in
stars. The redshift evolution of this scaling relation provides a
powerful test of galaxy formation models. Hierarchical models of
galaxy formation predict that the ratio of stellar mass to total mass
would be similar at all redshifts i.e. massive galaxies assembled
through mergers of smaller galaxies. The `classical' galaxy formation
models (e.g. monolithic collapse; Eggen, Lynden-Bell \& Sandage 1962)
predict a higher mass to light ratio at higher redshift, since the gas
is still being converted to stars within the fully-formed dark matter
halo. However, stellar population evolution might produce an offset in
the opposite sense (when based on observations at short rest-frame
wavelength), since at high redshift a stellar population of
identical mass is likely to be younger and therefore bluer than an
equivalent population at low redshift.

Investigation of the evolution of the rest frame $B$-band TF relation
at $z>0.5$ has to date produced discrepant results with observations
of only moderate luminosity evolution of $\la 0.4$ mag out to $z\sim1$
(Vogt et al. 1996, 1997) compared with claims of much stronger
evolution of $\sim2$ mag at $z\sim0.3$ (Simard \& Pritchet 1997; Rix
et al. 1997). Previous high-$z$ work has so far been limited to
long-slit spectroscopy, which has inherent problems - missing light
through slit losses and possible misalignment of the slit, potentially
leading to erroneous results for the rotation velocity if not
addressed properly. However, the relatively new technique of 3D
spectroscopy, used by Andersen \& Bershady (2003) at low redshift,
overcomes these difficulties. Using an integral field unit (IFU) we
can efficiently obtain a complete census of the spatially extended
line emission. We can potentially map disc kinematics at $z\sim 1$ out
to sufficiently large galactocentric radii to measure $v_{max}$
(without need for correction), as well as obtaining more detailed
information on the spatial distribution of emission line regions.

In this letter we present the first demonstration of near-IR integral
field spectroscopy of a $z\sim1$ field galaxy, using our new CIRPASS instrument
(Cambridge IR PAnoramic Survey Spectrograph, Parry et
al. 2000) on Gemini-South. By moving into the near-IR we were able to
study the star formation in the $z=0.8$ galaxy CFRS 22.1313 (Lilly et
al. 1995b) using the rest-optical H$\alpha$ $\lambda$ 6563 \AA\ emission
line, the same reliable tracer of star formation as used at low
redshift. We investigate the disc kinematics through the H$\alpha$
line and compare the results found here using an IFU to previous work
using traditional long-slit spectroscopy.

The layout of this letter is as follows. In section 2 we describe our
target galaxy, the CIRPASS IFU observations and archival HST
imaging. In section 3 we discuss the distribution of star formation
and the large-scale kinematics of the galaxy. Our conclusions are
presented in section 4. We assume $H_0=70$ km s$^{-1}$Mpc$^{-1}$,
$\Omega_M=0.3$ and $\Omega_{\Lambda}=0.7$ throughout this letter,
unless otherwise stated. An angular scale of $1$ arcsec corresponds to a
physical distance of 7.57 kpc at $z=0.819$ using the above cosmology.

\section{Observations and Data Reduction}

\subsection{CIRPASS Spectroscopy}

Our target is the disc galaxy CFRS 22.1313 (part of the Canada-France
Redshift Survey (CFRS) - Lilly et al. 1995a) and was specifically
chosen for IFU study as it is spatially extended and has an accurately
known redshift ($z=0.819$), for which H$\alpha$ appears between sky
lines. It also has strong [OII] $\lambda$ 3727\,\AA\ flux, which
should imply detectable H$\alpha$ emission (Kennicutt 1992), and benefits from deep HST
imaging from which the disc structural parameters may be derived.
Near-infrared integral field spectroscopy of CFRS 22.1313 was obtained
with CIRPASS on Gemini-South on the night of 2002 August 13. CIRPASS
is a fibre-fed spectrograph with a 490 lenslet array with a variable
lenslet scale and operates in the $J$- and $H$-band
($1$--$1.67\,\mu$m). Using the $0.25$ arcsec diameter lenslet scale
for these observations the array covered an area of $9.3$ by $2.9$
arcsec. The detector is a 1k$\times$1k Hawaii-I HgCdTe Rockwell
array. The CIRPASS observations comprised eight separate 1800\,s
integrations with a seeing of about $0.4$ arcsec. An eight-point
dither pattern was used to facilitate sky subtraction and the removal
of bad pixels. The spectra were taken using a 400 l mm$^{-1}$ grating
with resolving power $R=\frac{\lambda}{\Delta\lambda}\approx 3000$ and dispersion 2.28\,\AA\ pix$^{-1}$.
The wavelength coverage was $1$--$1.24\,\mu$m, targeting the redshifted
H$\alpha$ emission line, which appeared in a region free of OH sky emission
lines.

The data were reduced using the {\scriptsize CIRPASS IRAF} package\footnote{The
CIRPASS data reduction software is available from {\tt
http://www.ast.cam.ac.uk/$\sim$optics/cirpass/docs.html}}. For each
1800\,s integration the detector was read out non-destructively 20 times
at the start of the integration and at 600\,s, 1200\,s and 1800\,s. The 20
multiple array reads are averaged for each loop to reduce read noise
and adjacent loop averages are used to identify and reject cosmic
rays. Bias subtraction was also performed at this stage. Since the
target was stepped across different lenslets in the IFU for each of
the eight observations, a sky frame was created for each exposure from
the seven other integrations and this was used to perform first order
sky subtraction.  The spectra were then extracted using an optimal
extraction algorithm (Johnson, Dean \& Parry 2002) to account for crosstalk
between adjacent spectra on the detector and to trace the curvature of
the spectra from each fibre. The extracted spectra were flat-fielded
using a dome flat, in order to remove the variation in throughput
between fibres and the variation in sensitivity between pixels. The
dome flat had previously been read-averaged, bias-subtracted and
extracted.  The data were wavelength calibrated using an argon lamp
exposure, processed in the same way as the science data, again using
optimal extraction. We used 17 argon lines and a quadratic fit to the
dispersion, producing {\em rms} residuals of 0.2 \AA\ (0.1
pix). Residual sky-lines were removed by fitting a low-order
polynomial function to the wavelength-rectified data frames. Finally
3D (x,y,$\lambda$) data cubes were constructed for each integration
before combining the eight individual exposures using the known
telescope offsets. We observed two standard stars Hip257 and Hip106522
($J\approx9$ mag) to determine the flux calibration (these produced
consistent calibrations). The observations of the standard stars were
processed in the same way and used to flux calibrate the data cube.

\begin{figure}
\resizebox{0.98\columnwidth}{!}{\includegraphics{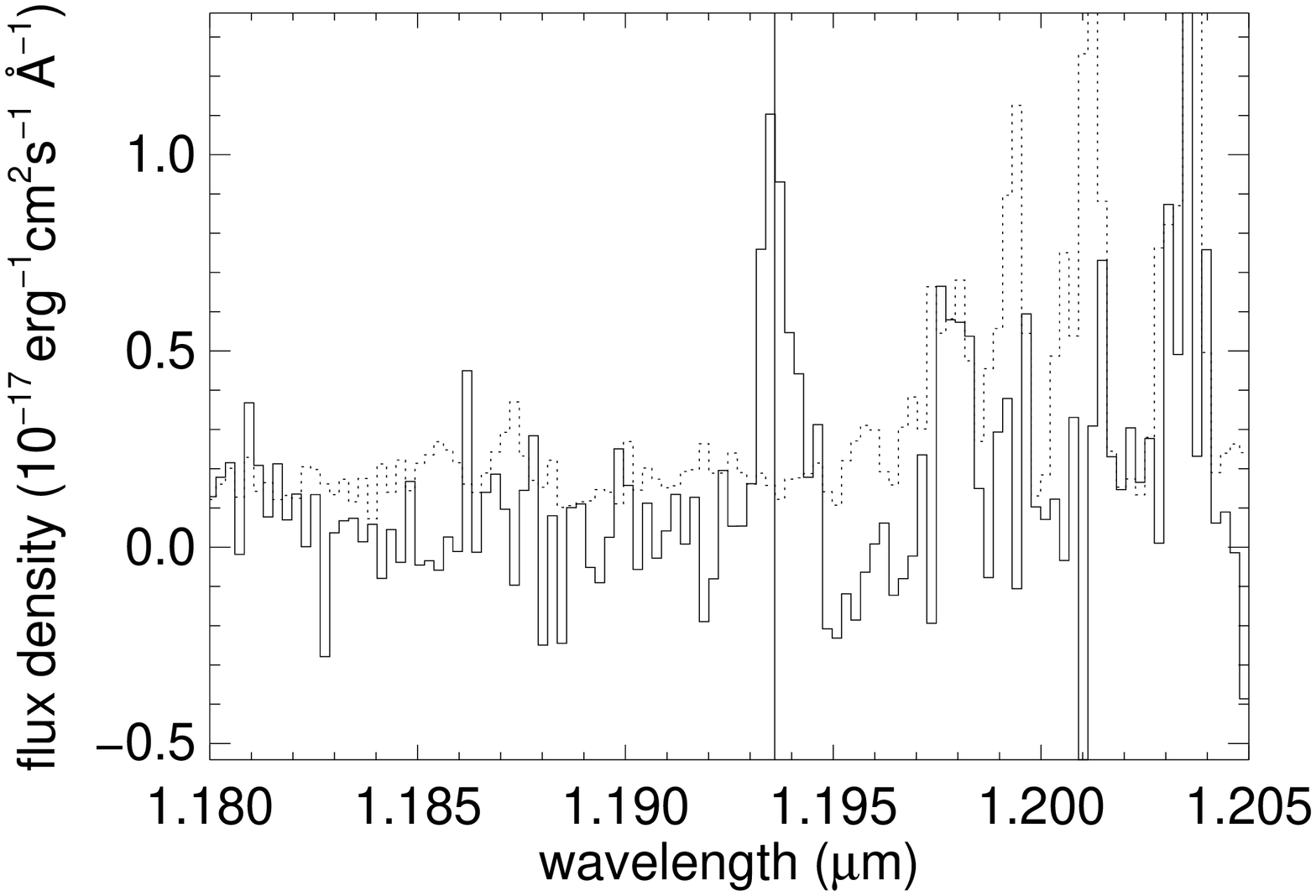}}
\resizebox{0.98\columnwidth}{!}{\includegraphics{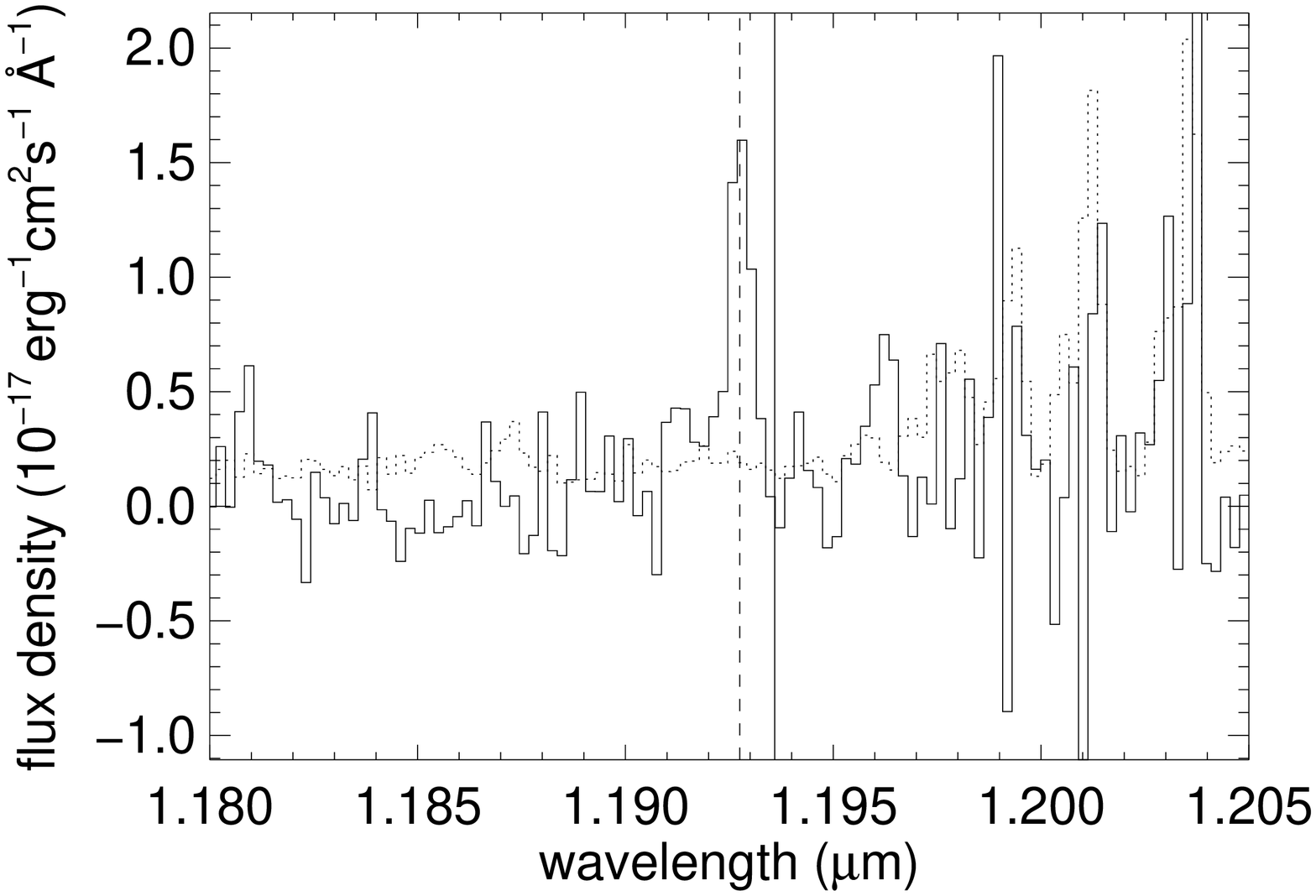}}
\caption{The upper panel shows the spectrum for knot A  (solid line), the lower spectrum is for knot B (solid line), both for extraction apertures of $\sim 1$ arcsec $\times 1$ arcsec. The centre of the H$\alpha$ emission for knot A (solid vertical line) is clearly offset from the central wavelength for knot B (dashed line). The dotted spectrum shows the sky emission. Sky subtraction residuals are still visible, although by selection the H$\alpha$ line for this galaxy lies away from the sky lines.}
\label{fig:spectra}
\end{figure}

\subsection{HST Photometry}
HST WFPC2 imaging was obtained from the archive (programme GO5996, the
CFRS survey). There were five $I$-band F814W exposures with a total
exposure time of 6700\,s. We re-reduced the data from the archive
starting with the pipeline processed (flat-fielded and
dark-subtracted) individual exposures. These were then combined with
integer pixel shifts determined from the world coordinate system and
with cosmic ray rejection at the 3$\sigma$ level.

The {\scriptsize GIM2D} software package (Simard 1998; Simard et
al. 1999) was used to determine an accurate inclination, flux and
scale length for the galactic disc in the HST F814W image. The
approach of {\scriptsize GIM2D} is to convolve idealized
two-dimensional galaxy models with the instrumental point spread
function (PSF) and to subtract this from the original two-dimensional
image: the best-fitting model is found by altering the model
parameters to yield the smallest residuals from the real image.

Synthetic instrumental PSFs were created by {\scriptsize TinyTim v5.0}
(Krist 1995), which has the advantage that they are noiseless, can be
sub-sampled and are created at the required location on the array. We
produced PSFs sub-pixelated by a factor of 5 for WFPC\,2\,/\,WF\,2
(F814W).  We considered only an exponential disc profile ($\log
I\propto {1/r}$): as we are working in the rest-frame $B$-band at
$z=0.8$, the red bulge should produce negligible deviations from the
fit of the exponential disc to the WFPC\,2 F814W $I$-band data. We
excluded the two H{\scriptsize~II} regions which were bright in the
rest-$B$ (section~\ref{sec:HIIregions}) from the {\scriptsize GIM2D}
fit to the surface brightness profile, by masking regions of width 0.5
arcsec centred on the two knots. We experimented with different mask
sizes and found this did not significantly affect the disc parameters
and these errors are included in the uncertainty in the rotation
velocity. 

The galaxy centre, position angle on the sky, disc scale length,
inclination and total flux were free parameters in the model.  The
best-fitting disc model from {\scriptsize GIM2D} had a half-light
radius of $1.25$ arcsec (10 kpc): the disc itself had a position angle
on the sky of 74\,$\pm\,2^{\circ}$, the inclination to the line of
sight was measured to be 78\,$\pm\,2^{\circ}$ (sin $i =0.98$). The
exponential disc scale length was 0.75\,$\pm\,0.05$ arcsec
(5.7\,$\pm\, 0.4$ kpc). NICMOS NIC2 imaging for this galaxy also
exists. We have examined this and find that the centre determined from
the $H$-band imaging (which traces the older starlight) agrees with
the centre presented here.

Using SExtractor (Bertin \& Arnouts 1996) the galaxy was found to have
$I_{\rm Vega}=21.7$ with the H{\scriptsize~II} regions contributing
$\sim$ 25 per cent of the total flux. Galactic extinction was
determined using reddening $E(B-V)=$0.05 from Burstein \& Heiles
(1982). A correction of 0.88\,mag was made for the
inclination-dependent internal extinction, including the residual
absorption of a face-on galaxy (0.27mag), using the prescription of
Tully \& Fouqu\'{e} (1985) with an optical depth $\tau=$0.55. The
magnitudes were adjusted by a small k-correction (a dimming of 0.02mag
for a Sbc spiral template spectrum from Coleman, Wu \& Weedman 1980) to
match the observed $I$-band to the rest-frame $B$-band.

\begin{figure*}
\resizebox{0.66\columnwidth}{!}{\includegraphics{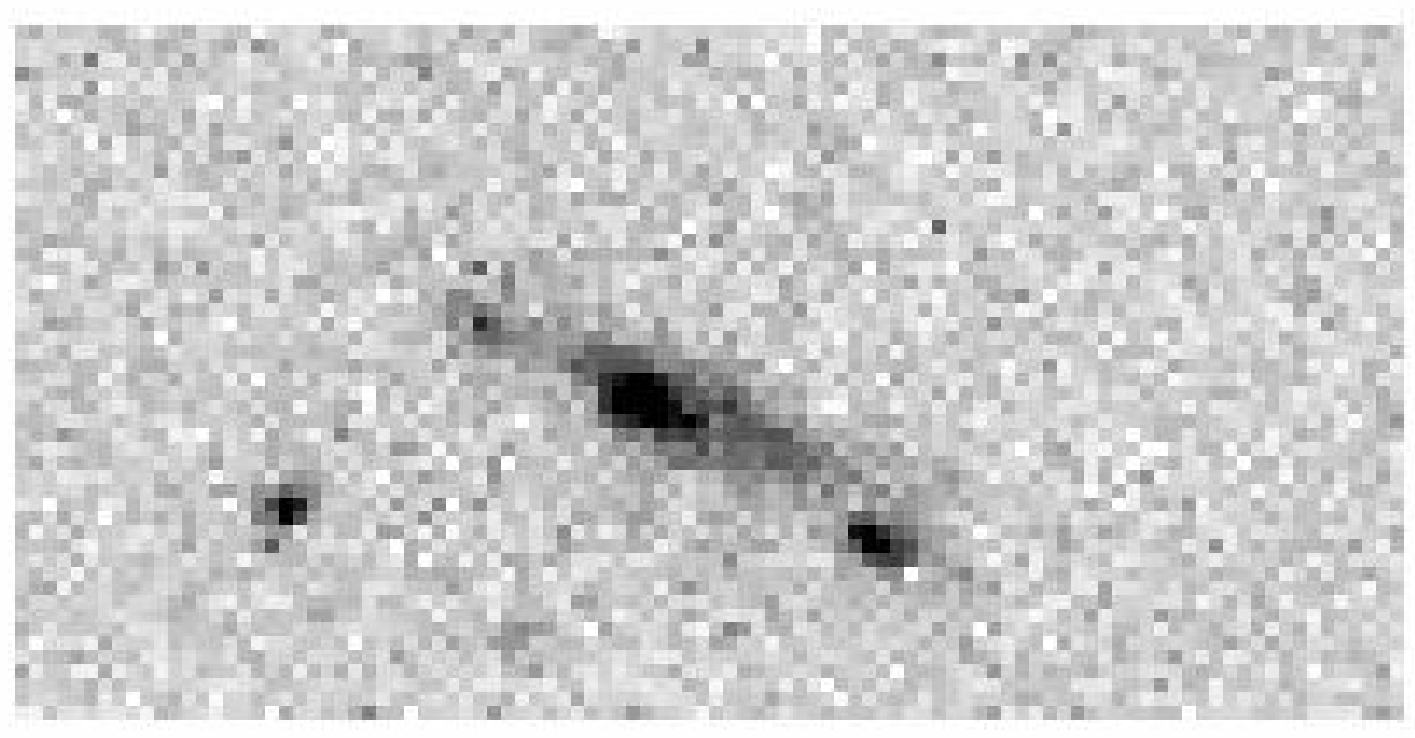}}
\resizebox{0.66\columnwidth}{!}{\includegraphics{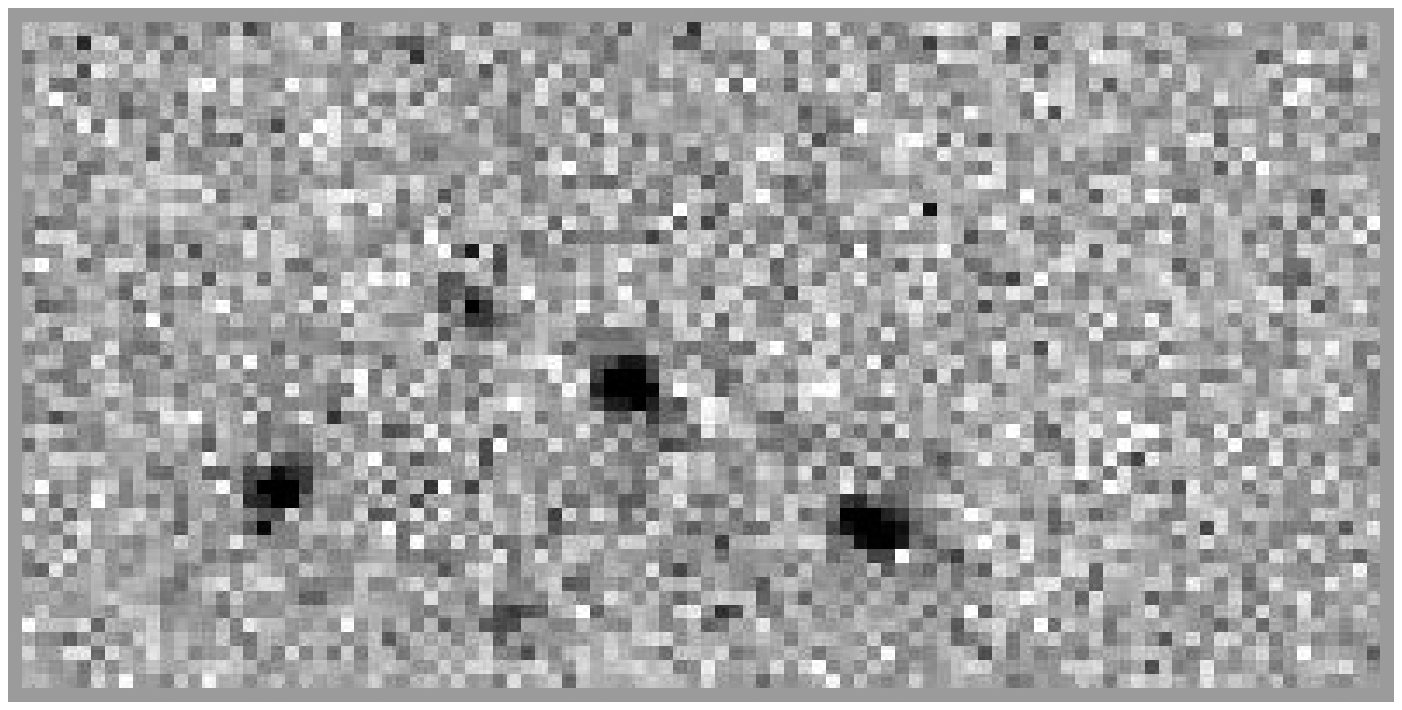}}
\resizebox{0.66\columnwidth}{!}{\includegraphics{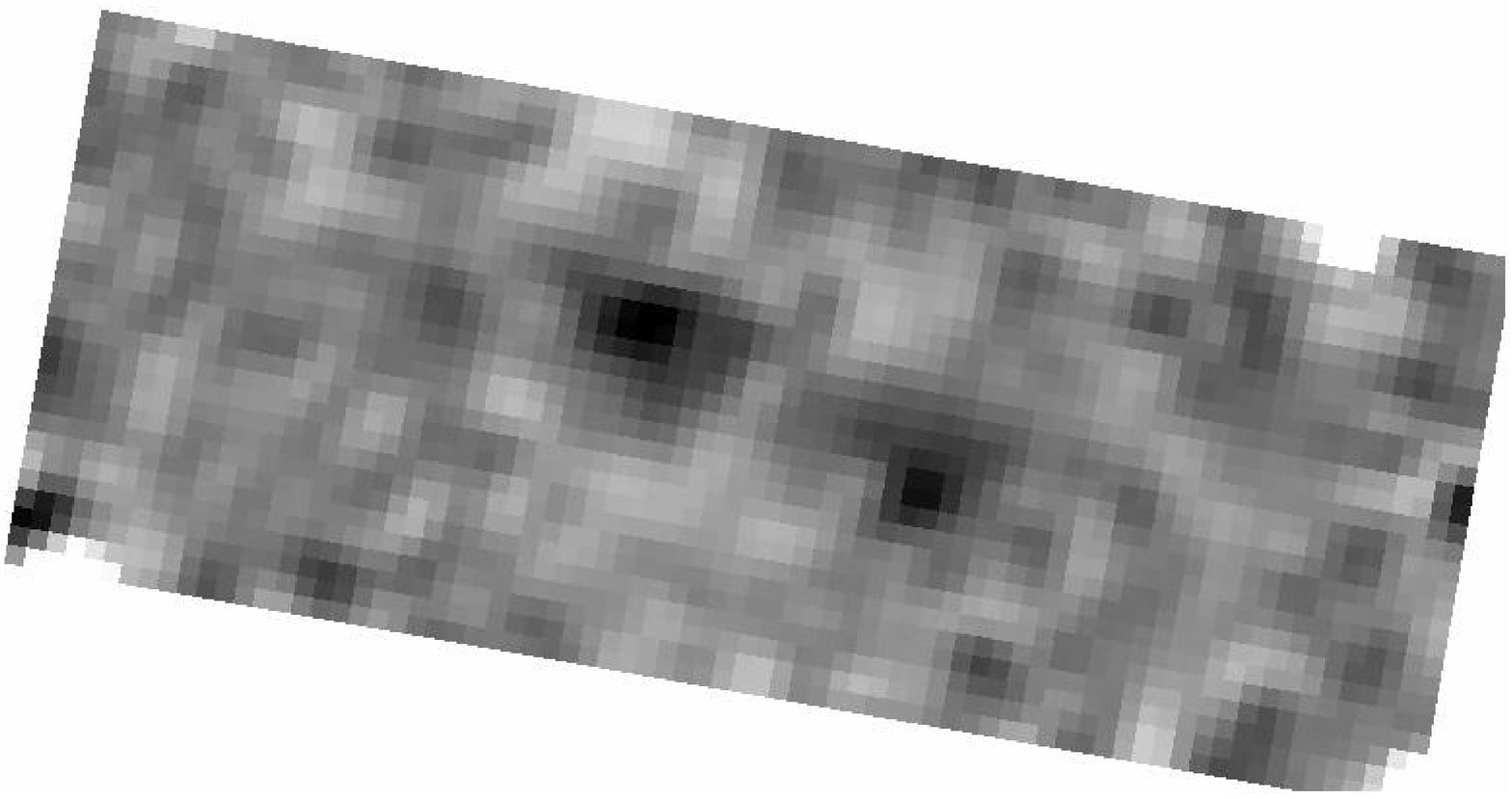}}
\put(-350,80){\line(1,0){15}}
\put(-358,73){1 arcsec}
\put(-388,67){\vector(-1,-1){15}}
\put(-388,69){A}
\put(-360,50){\vector(-1,-1){15}}
\put(-360,52){B}
\put(-408,38){\line(1,1){10}}
\put(-398,38){\line(-1,1){10}}
\put(-82,35){\line(1,1){10}}
\put(-72,35){\line(-1,1){10}}
\put(-470,80){\line(1,0){19.8}} 
\put(-472,73){10 kpc}
\put(-335,20){\rotatebox{-9}{\vector(-1,0){20}}}
\put(-335,20){\rotatebox{-9}{\vector(0,1){20}}}
\put(-362,20){E}
\put(-335,43){N}
\caption{Left: the HST F814W image (corresponding to the rest frame $B$-band at $z=0.8$). The field has been trimmed to $6 \times 11$ arcsec. The two prominent H{\scriptsize~II} regions are clearly visible against the extended highly inclined galactic disc. Middle: the HST F814W image with the best-fitting disc model from {\scriptsize GIM2D} subtracted. Right: these regions are seen in our CIRPASS near-infrared spectrum. We show a 2D representation of our 3D data cube, collapsed along the spectral direction and showing the $\sim$15 \AA\ around the H$\alpha$ emission. The spatial scale and orientation are identical to that of the HST WFPC2 image.}
\label{fig:IFU}
\end{figure*}

\section{Discussion}

\subsection{Distribution of star formation}
\label{sec:HIIregions}

In this paper we present a map of the distribution of star-forming
regions within this $z=0.819$ galaxy. Unlike long-slit approaches this
is a complete census of the spatially extended star formation. As can
be seen from Fig. \ref{fig:IFU} the H$\alpha$ emission is concentrated
in just two star-forming regions and these are spatially coincident
with the `knots' seen in the HST $I$-band image (which corresponds to
the rest frame $B$).  From the HST image both of these `knot'-like
H{\scriptsize~II} regions are spatially resolved and noticeably
elongated in the plane of the galactic disc. Knot A is near coincident
with the projected centre of the galactic disc and extends over $\sim$
2.7 $\times$ 2.4 kpc FWHM (after deconvolution with the WFPC2
PSF). Knot B is about $2$ arcsec from the centre and is $\sim$ 2.8
$\times$ 1.8 kpc FWHM (deconvolved). Their properties are tabulated in
Table \ref{table:HII}. In determining the rest frame $B$-band
luminosity of the star-forming knots we have first subtracted off the
best-fitting galactic disc model provided by {\scriptsize GIM2D}
(section 2.2). The surface brightness within the half-light radius of
the star-forming knots in rest frame $B$ correspond to $10^9$
M$_{\odot}$ kpc$^{-2}$ within the range given in Meurer et al. (1997)
for starburst galaxies at $0<z<3.5$. We note that knot B (the
H{\scriptsize~II} region with the largest projected separation from
the centre) lies exactly on the bolometric surface brightness versus
angular frequency relation given in Meurer et al. (1997), based on the
sample of Lehnert \& Heckman (1996).  We measure H$\alpha$ at the
greater than 5$\sigma$ level for each knot with fluxes of
$\sim10^{-16}$ ergs cm$^{-2}$ s$^{-1}$ each
(Fig. \ref{fig:spectra}). We note our total H$\alpha$ luminosity of
$2.4 \times 10^{-16}$ erg s$^{-1}$ cm$^{-2}$ is slightly less than the value of
$4 \times 10^{-16}$ obtained by Glazebrook et al. (2001) in their CGS4
UKIRT spectroscopy. In the absence of reddening our H$\alpha$ fluxes
would correspond to star formation rates of $\sim3\,$M$_{\odot}\,{\rm
yr}^{-1}$ per knot, where we have adopted the prescription of
Kennicutt (1998), applicable for a Salpeter-like initial mass function
(IMF). We note that the rest frame $B$ flux densities (Table
\ref{table:HII}) would underestimate these star formation rates (for
the same IMF) by a factor of $\sim2$, which we attribute to the
differential dust reddening between 4400 and 6563 \AA\ or to the fact
that the $B$-band flux, unlike H$\alpha$, is not a direct measurement
of current star formation rate.  We spectrally resolve the line widths
within the individual H{\scriptsize~II} regions
(Fig. \ref{fig:spectra}). After deconvolution with the CIRPASS
instrumental line width we measure $\sigma_v$ of 44\,$\pm$\,15 and
74\,$\pm$\,10 km s$^{-1}$.

\begin{table*}
\caption{Properties of star-forming regions} 

\begin{tabular}{cccccccccc}
\hline
{}& d & $I$ & F$^B_\nu$ & F(H$\alpha$) & $\lambda_{cen}$ & FWHM & v & $\sigma_v$ & SFR(H$\alpha$)\\
{} & (kpc)& (mag) & ($10^{28}$erg s$^{-1}$ Hz$^{-1}$)& ($10^{-17}$erg s$^{-1}$ cm$^{-2}$)&(\AA)& (\AA) & (km s$^{-1}$)& (km s$^{-1}$)& (M$_\odot$ yr$^{-1}$)\\
{}&(1)&(2)&(3)&(4)&(5)&(6)&(7)&(8)&(9)\\
\hline
knot A &{2.8\,$\pm$\,1.0}&{24.11}&{1.00}&{9.4}&{11935.9\,$\pm$\,0.4}& 7.9\,$\pm$\,1.0 &{40\,$\pm$\,20}&{74\,$\pm$\,10}&{2.7}\\
knot B &{12.6\,$\pm$\,1.0}&{24.10}&{1.01}&{14.7}&{11927.5\,$\pm$\,0.4}& 5.8\,$\pm$\,1.0 &{180\,$\pm$\,20}&{44\,$\pm$\,15}&{4.2}\\
\hline
\end{tabular}

The columns are as follows:
(1) projected distance of star-forming knot from centre of galactic disc,
(2) total magnitude (MAG\_AUTO from SExtractor, Bertin \& Arnouts 1996) in model-subtracted $I$-band image,
(3) rest-$B$ flux density from model-subtracted $B$-band image,
(4) H$\alpha$ line flux,
(5) central wavelength of H$\alpha$ emission,
(6) FWHM of H$\alpha$ line,
(7) velocity shift with respect to centre, 
(8) velocity dispersion $\sigma_v$ (deconvolved),
(9) star formation rate from H$\alpha$.

\label{table:HII}
\end{table*}

\subsection{Large scale kinematics}

Using the H$\alpha$ emission line, we measure a velocity separation
$v_{sep} =$ 220\,$\pm$\,10 km s$^{-1}$ for knots A and B
(Table~\ref{table:HII}). However, we do not measure a rotation curve
for this galaxy: we simply have two discrete line emission
regions. Coupling our CIRPASS IFU 3D spectrum with the HST images
(specifically the location of the H{\scriptsize~II} regions and the
best fit to the centre of the galactic disc) we can set a robust lower
limit of 180\,$\pm$\,20 km s$^{-1}$ to $v_{max}$ (where $v_{max}$ is
defined as half the total velocity shift across the galaxy). Of course
as we see no evidence for  a `turnover' in the rotation curve
$v_{max}$ could be significantly higher. We note that our lower limit
on $v_{max}$ is significantly greater than the value of 120\,$\pm$\,10
km s$^{-1}$ presented by Barden et al. (2003) for the same galaxy
based on their ISAAC/VLT near-IR spectroscopy. This is $\sim
\frac{1}{2} v_{sep}$ and would be consistent with our measurement of
the redshift difference between the H{\scriptsize~II} regions if it
was assumed that the centre of rotation was the geometric midpoint of
the two star-forming knots. However, the HST image does not support
this, with knot B at far greater projected distance from the disc
centre. The systemic redshift is likely to be close to the H$\alpha$
redshift of knot A ($z=0.8187$ cf. $z=0.8191$ from the CFRS catalogue,
Lilly et al. 1995b).

A simple estimate of the mass of the galaxy, $M=V^2R/G$, yields a
value of $0.9 \times 10^{11}$ M$_{\odot}$ within 12 kpc.

\subsection{The Tully-Fisher Relation at $z\sim1$}

\begin{figure}
\resizebox{0.98\columnwidth}{!}{\rotatebox{90}{\includegraphics{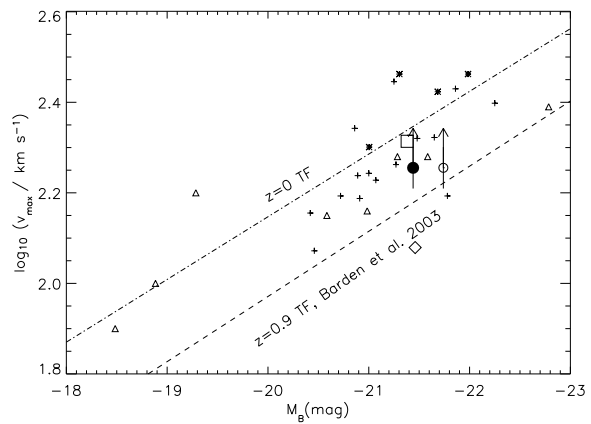}}}
\caption{The rest frame $B$-band TF relation showing the position of
CFRS 22.1313 with $B$-band flux from the star-forming regions excluded
(filled circle) and included (open circle), compared with high
redshift ($0.6<z<1$) data from B\"{o}hm et al. 2004 (crosses, high
quality data only), Vogt et al. 1996 (stars), the field galaxy sample
from Milvang-Jensen et al. 2003 (triangles) and the Barden et al. 2003 data
point for CFRS 22.1313 (diamond). IFU data for a $z=1$ lensed arc from Swinbank
et al. 2003 is also plotted (open square). The dashed line shows the
high redshift TF relation of Barden et al. (2003). The dot-dash line
shows the local TF relation of Tully and Pierce (2000).}
\label{fig:tfplot}
\end{figure}

The TF relation is a fundamental scaling law, the evolution of which
is critical to our understanding of mass assembly and star formation
in disc galaxies.  Evolution of the TF relation at $z>0.5$ has been
investigated with varied results. Vogt et al. (1996, 1997) find only
moderate luminosity evolution ($\Delta M \la 0.4$) out to $z\sim1$
while Barden et al. (2003), Milvang-Jensen et al. (2003) and B\"{o}hm
et al. (2004) detect a brightening of $\sim1$ mag in the rest-frame
$B$-band. Much stronger evolution is suggested by the study of Simard
\& Pritchet (1997), which finds disc brightening of $\sim2$ mag at
lower redshift ($z\sim0.35$) and Rix et al. (1997), which derives a
brightening of 1.5 mag at $<z>=0.25$ using the {\scriptsize [OII]}
line. The degree of evolution observed for a particular sample is
likely to depend strongly on the selection method.

In Fig.~\ref{fig:tfplot} we use our measured lower limit on $v_{max}$
to place CFRS 22.1313 on the TF relation and compare with the TF
relation at the current epoch (Tully \& Pierce 2000). We also plot the
estimated TF relation at redshifts $0.6<z<1$ ($\Delta z=0.2$ either
side of the redshift of CFRS 22.1313) derived by other groups using
long-slit spectroscopy. Using the total HST F814W magnitude for the
galaxy of $I_{\rm Vega}=21.7$ we obtain a lower limit on the offset of
this one galaxy from the local rest $B$-band TF relation consistent
with disc brightening of no more than $\sim$1 mag at $z=0.8$ and 
entirely consistent with no evolution.

\section{Conclusion}
Our main results may be summarised as follows:
\begin{enumerate}
\item We have presented the first near-IR integral field spectrum of a
$z\sim1$ field galaxy. We have used our CIRPASS integral field
spectrograph to map the spatial and velocity distribution of H$\alpha$
in the $z=0.819$ galaxy CFRS 22.1313.
\item The H$\alpha$ emission originates from two giant
H{\scriptsize~II} regions offset by 220\,$\pm$\,10 km s$^{-1}$ : we do not
observe a full rotation curve for this galaxy.
\item Archival HST imaging shows a clear disc morphology, along with
two bright star-forming knots coincident with the H$\alpha$ emission and
accounting for $\sim$ 25 per cent of the total $I$-band flux (the rest frame
$B$-band).
\item We have coupled the 3D spectroscopy with the best-fitting model
to the underlying disc in the HST F814W image, and used this to derive
a lower limit to the rotation velocity of 180\,$\pm$\,20 km
s$^{-1}$. The lower limit on the offset of this one galaxy from the
local rest $B$-band TF relation is consistent with disc brightening of
no more than $\sim$1 mag at $z=0.8$ and entirely consistent with no
evolution.
\item Integral field spectroscopy eliminates many of the systematic
uncertainties inherent in long-slit studies. A larger sample of IFU data is
required to accurately determine the evolution of the TF relation out
to $z\sim1$.
\end{enumerate}

\subsection*{ACKNOWLEDGMENTS}
This paper is partially based on observations obtained at the Gemini
Observatory, which is operated by the Association of Universities for
Research in Astronomy, Inc. (AURA), under a cooperative agreement with
the U.S. National Science Foundation (NSF) on behalf of the Gemini
partnership: the Particle Physics and Astronomy Research Council
(PPARC, UK), the NSF (USA), the National Research Council (Canada),
CONICYT (Chile), the Australian Research Council (Australia), CNPq
(Brazil) and CONICET (Argentina). We are grateful to Matt Mountain for
the Director's discretionary time to demonstrate the scientific
potential of integral field units (the PI s of this demonstration
science programme are Andrew Bunker, Gerry Gilmore and Roger
Davies). We thank the Gemini Board and the Gemini Science Committee
for the opportunity to commission CIRPASS on the Gemini-South
telescope as a visitor instrument. We thank Phil Puxley, Jean
Ren\'{e}-Roy, Doug Simons, Bryan Miller, Tom Hayward, Bernadette
Rodgers, Gelys Trancho, Marie-Claire Hainaut-Rouelle and James Turner
for the excellent support received. CIRPASS was built by the
instrumentation group of the Institute of Astronomy in Cambridge,
UK. We warmly thank the Raymond and Beverly Sackler Foundation and
PPARC for funding this project. Andrew Dean, Anamparambu Ramaprakash
and Anthony Horton all assisted with the observations in Chile and we
are indebted to Dave King, Jim Pritchard and Steve Medlen for
contributing their instrument expertise. The optimal extraction
software for this 3D fibre spectroscopy was written by Rachel Johnson
and Andrew Dean. This research is also partially based on observations
with the NASA/ESA Hubble Space Telescope, obtained at the Space
Telescope Science Institute (STScI), which is operated by AURA under
NASA contract NAS 5-26555. This research was supported by NSF grant
NSF-0123690 via the ADVANCE Institutional Transformation Program at
NMSU. JKS acknowledges a PPARC studentship supporting this study. We
thank the anonymous referee for helpful comments on this manuscript.


\begin{thebibliography}{}
\bibitem [\protect\citename{Andersen \& Bershady} 2003]{an03} 
Andersen D.R., Bershady M.A., 2003, ApJ, 599, L79

\bibitem [\protect\citename{Bertin \& Arnouts} 1996]{be96}
Bertin E., Arnouts S.,1996, A\&AS, 117, 393

\bibitem [\protect\citename{Barden et al.\ } 2003]{ba03} 
Barden M., Lehnert M.D., Tacconi L., Genzel R., White S., Franceschini A., preprint (astro-ph/0302392)

\bibitem [\protect\citename{Burstein \& Heiles} 1982]{bu82} 
Burstein D., Heiles C., 1982, AJ, 87, 1165

\bibitem  [\protect\citename{B\"{o}hm et al.\ } 2004]{bo04} 
B\"{o}hm A. et al., 2004, A\&A, 420, 97

\bibitem  [\protect\citename{Coleman, Weedman \& Wu} 1980]{c080}
Coleman G.D., Wu C.-C., Weedman D.W., 1980, ApJS, 43, 393

\bibitem  [\protect\citename{Eggen et al.\ } 1962]{eg62} 
Eggen O.J., Lynden-Bell D., Sandage A.R., 1962, ApJ, 136, 748

\bibitem  [\protect\citename{Glazebrook et al.\ } 1999]{gl99} 
Glazebrook K., Blake C., Economou F., Lilly S., Colless M., 1999, MNRAS, 306, 843

\bibitem  [\protect\citename{Johnson et al.\ } 2002]{jo02} 
Johnson R.A., Dean A.J., Parry I.R., 2002, in Rosado M., Binette L., Arias L.,eds, ASP Conf. Ser. Vol. 282, Galaxies: The Third Dimension.Astron. Soc. Pac., San Francisco, p.531

\bibitem  [\protect\citename{Kennicutt\ } 1992]{ke92} 
Kennicutt R.C., 1992, ApJ, 388, 310


\bibitem  [\protect\citename{Kennicutt\ } 1998]{ke98} 
Kennicutt R.C., 1998, ARA\&A, 36, 189

\bibitem  [\protect\citename{Krist et al.\ } 1995]{kr95} 
Krist J., 1995, in Shaw R.A., Payne H.E., Hayes J.J.E., eds, ASP Conf. Ser. Vol. 77, Astronomical Data Analysis Software and Systems IV.Astron. Soc. Pac., San Francisco, p.349

\bibitem  [\protect\citename{Lehnert \& Heckman} 1996]{le96} 
Lehnert M., Heckman T.M., 1996, ApJ, 472, 546

\bibitem  [\protect\citename{Lilly et al.\ } 1995]{li95a} 
Lilly S.J., Le Fevre O., Crampton D., Hammer F., Tresse L., 1995a, ApJ, 455, 50

\bibitem  [\protect\citename{Lilly et al.\ } 1995]{li95b} 
Lilly S.J., Hammer F., Le Fevre O., Crampton D., 1995b, ApJ, 455, 75

\bibitem  [\protect\citename{Meurer et al.\ } 1997]{me97} 
Meurer G.R., Heckman T.M., Lehnert M.D., Leitherer C., Lowenthal J., 1997, AJ, 114, 54 

\bibitem  [\protect\citename{Milvang-Jensen et al.\ } 2003]{mi03} 
Milvang-Jensen B., Aragon-Salamanca A., Hau G.K.T., Jorgenson I., Hjorth J., 2003, MNRAS, 339, 1

\bibitem  [\protect\citename{Parry et al.\ } 2000]{pa00} 
Parry I.R. et al., 2000, SPIE, 4008, 1193

\bibitem  [\protect\citename{Rix et al.\ } 1997]{ri97} 
Rix H.-W., Guhathakurta P., Colless M., Ing K., 1997, MNRAS, 285, 779   

\bibitem  [\protect\citename{Simard} 1998]{si98} 
Simard L., 1998, in Albrecht R., Hook R.N., Bushouse H.A., eds, ASP Conf. Ser. Vol. 145, Astronomical Data Analysis Software and Systems VII.Astron. Soc. Pac., San Francisco, p.108

\bibitem  [\protect\citename{Simard} 1999]{si99} 
Simard L. et al., 1999, ApJ, 519, 563

\bibitem  [\protect\citename{Simard \& Pritchet} 1998]{sipr98} 
Simard L., Pritchet C.J., 1998, ApJ, 505, 96

\bibitem  [\protect\citename{Swinbank et al.\ } 2003]{sw03} 
Swinbank A.M. et al., 2003, ApJ, 598, 162

\bibitem  [\protect\citename{Tully \& Fisher} 1977]{tf77} 
Tully R.B., Fisher J.R., 1977, A\&A, 54, 661

\bibitem  [\protect\citename{Tully \& Fouque} 1985]{tf85} 
Tully R.B., Fouqu\'{e} P., 1985, ApJS, 58, 67

\bibitem  [\protect\citename{Tully \& Pierce} 2000]{tp00} 
Tully R.B., Pierce M.J., 2000, Apj, 533, 744

\bibitem  [\protect\citename{Vogt et al.\ } 1996]{vo96} 
Vogt N.P., Forbes D.A., Phillips A.C., Gronwall C., Faber S.M, Illingworth G.D., Koo D.C., 1996, ApJ, 465L, 15

\bibitem  [\protect\citename{Vogt et al.\ } 1997]{vo97} 
Vogt N.P. et al., 1997, ApJ, 479, 121 
\end{thebibliography}
\end{document}